\documentstyle[aps,prc,epsfig]{revtex}
\begin{document}
\def\beq{\begin{equation}}
\def\eeq{\end{equation}}
\def\beqa{\begin{eqnarray}}
\def\eeqa{\end{eqnarray}}
\def\MeV{\nobreak\,\mbox{MeV}}
\def\GeV{\nobreak\,\mbox{GeV}}
\def\keV{\nobreak\,\mbox{keV}}
\def\fm{\nobreak\,\mbox{fm}}
\def\Tr{\mbox{ Tr }}
\def\pli{p^\prime}
\def\mli{{M^\prime}^2}
\def\mm{{M_m}^2}

\title{\sc On the $x_F$ distribution of  $J/\psi$'s produced in heavy ion collisions} 
\author { F.O. Dur\~aes,  F.S. Navarra  and M. Nielsen\\
\vspace{0.3cm}
{\it Instituto de F\'{\i}sica, Universidade de S\~{a}o Paulo, } \\
{\it C.P. 66318,  05389-970 S\~{a}o Paulo, SP, Brazil}}
\maketitle
\vspace{1cm}

\begin{abstract}
Thermal production of $J/\psi$ within quark gluon plasma is reconsidered. We show that if screening 
effects are not strong enough, the ``in-plasma born'' $J/\psi$'s would show up as a peak in the Feynman
momentum distribution at $x_F=0$.
\end{abstract}

\vspace{1cm}

\section{Introduction}

Over the past fifteen years charmonium suppression has been considered as one of the best signatures of 
quark gluon plasma (QGP) formation. Recently this belief was questioned by some works. 
$J/\psi$ suppression in relativistic  heavy ion collisions is based on the following simple argument,  presented 
in the original work by  Matsui and Satz \cite{msatz}.  Lattice simulations show that the heavy quark-antiquark 
potential becomes screened at high enough temperatures. As a consequence, when 
the range of the potential becomes smaller than the $J/\psi$  Bohr radius the bound state can no longer be formed. 
On the other hand,  detailed simulations \cite{thews} 
of a population of $c\bar{c}$ pairs traversing the plasma suggested that, given the 
large number of such pairs, the recombination effect of the pairs into  charmonium Coulomb bound states is 
non-negligible and might even  lead to an enhancement of $J/\psi$ production. This conclusion 
received support from  the calculations of \cite{rapp}, where a two component model for $J/\psi$ production was
proposed.

Taking the existing calculations seriously,  
it is no longer clear that an overall suppression of the number of $J/\psi$'s
 will be a signature of QGP. A more detailed analysis is 
required and a more complex pattern can emerge. In particular, we might have suppression in some regions of the 
phase space and enhancement in others. Indeed, this is the result of the analysis presented in \cite{huf}. In that 
work the authors study the fate of $c\bar{c}$ pairs produced in the early stage of heavy ion collisions, 
comparing the case where they have to traverse QGP with the case where they have to traverse ordinary nuclear matter. 
One of the merits of that paper is to emphasize that very interesting information can be extracted from the 
scaled momentum ($x_F$) distribution of the produced $J/\psi$'s.

Motivated by these observations, in this work we address the $J/\psi$ Feynman momentum distributions. In some 
aspects we follow refs. \cite{rapp} and  \cite{huf} with one important difference: we include $J/\psi$  production 
{\it within the plasma}. This is usually neglected because the number of $c \bar c$ pairs produced in the 
plasma is believed to be small.  However, as it will be discussed in section III, a closer look into the existing 
estimates shows discrepancies of two orders of magnitude.   
In \cite{wang}, for example, it was estimated to be of the order of $1 \%$ of the total number of
charm quark pairs. In \cite{levai}, with the inclusion of thermal parton masses this fraction was estimated to 
reach $20-30 \%$.  Finally, in \cite{lmw} this number could, in some cases, 
be equal to the number of directly produced pairs. 
No systematic effort was made to reconcile these different 
estimates mainly because the screening mechanism was believed to destroy all bound states. Therefore  in-plasma 
charm production has been an issue of open charm physics whereas the main focus has always been on hidden charm 
production and suppression.

Although well established in several 
lattice calculations \cite{karsch}, the  color screening mechanism may be not so effective 
in real collisions. The existing lattice results are valid for an infinite mass, static,  charge-anticharge 
pair and we know that this is not very close to the real situation, where charges are not so heavy and are moving.  
Recently,  a calculation of the screening effect in moving charges \cite{sidi} showed that, in this case the 
screening is not so strong.  With the improvements of lattice gauge determinations \cite{karsch} 
of the temperature-dependent quark-antiquark potential new calculations of the quarkonium spectrum in QGP were 
performed. In  \cite{digal} it was found that at up to temperatures of the order of $1.10 \,\, T_c$ the $J/\psi$ 
can survive as a bound state in the plasma. At higher temperatures the potential is too shallow to hold a bound 
charmonium state.

We shall assume that screening is not so strong and,  in a narrow temperature window just above the phase 
transition,  will allow for charmonium  (Coulomb) bound states  which  ``survive''  from QGP.  More precisely, we 
will allow up to one in hundred bound states to survive in the region where they should be screened and destroyed.
As it will be seen, these ``survivors'' 
tend to escape with $x_F \simeq 0$, thus giving us a new kind of QGP signal. Moreover, this peak at low $x_F$ will
fill the dip predicted in \cite{huf}.

We will consider $J/\psi$ production in nucleus-nucleus collisions by three different mechanisms:
i) direct (primordial QCD parton fusion)  ii) thermal (statistical coalescence at hadronization)  and 
iii) QGP (in-plasma parton fusion).  As already discussed in \cite{rapp} i) and ii) are just two 
extreme cases. In realistic  simulations, for a fixed number of $c \bar c$ pairs and of $J/\psi$'s, these last 
are gradually destroyed, giving origin to the ``regenerated'' component ii). In the particular calculation 
presented  in \cite{rapp}, for central collisions and for high enough energies a near to complete replacement of 
the initial by the final thermal $J/\psi$'s  was found.
Component iii) was considered long time ago \cite{cleynmans}, before the work of Matsui and Satz, and then 
it was left in oblivion.  Once produced the charmonium state will suffer interactions with the partons 
in the plasma, or with hadrons in hot and dense medium and in a later stage with comoving hadrons \cite{nos1,nos2}.

Before performing detailed calculations one could try to guess the shape of the outcoming $J/\psi$'s $x_F$ 
distribution in the three mechanisms mentioned above.  QCD production has been studied in the past and the 
naive extrapolation of the spectra measured in $p-p$ and $p-A$ collisions would lead us to conclude that the 
$J/\psi$'s produced in i) should follow a distribution of the type \cite{grib}:
\begin{center}
\beq
\simeq \,\,\,\,  (1-x_F)^c
\label{qcd1}
\eeq
\end{center}
where $c \simeq 5 - 6$. As it will be discussed in the next section, 
perturbative QCD (pQCD) calculations 
within the framework of the color evaporation model (CEM) confirmed this 
behavior in leading order and next-to-leading order 
\cite{ramona}. No explicit calculation of this $x_F$ distribution was 
performed with the color octet model (COM), but a recent study of
asymmetries in open  charm production \cite{bra} suggests that a similar 
smooth behavior at $x_F \simeq 0$ might be found for the $J/\psi$ differential
cross section. Of course, this has to be verified. 
Thermal production follows a Bose-Einstein distribution 
\cite{munz}, which integrated in $p_T^2$ gives:
\begin{center}
\beq
\simeq  \,\,\,\,exp \left[ -\frac{1}{T} \,\left( \sqrt{x_F^2\,s/4 + m_{J/\psi}^2} \,\right) \right] \,\,
\left[1 + \frac{1}{T} \, \sqrt{x_F^2\,s/4 + m_{J/\psi}^2} \,\right]
\label{bose}
\eeq
\end{center}
where $x_F$ and $\sqrt{s}$ are the Feynman momentum and the c.m.s. nucleon-nucleon invariant energy.  Comparing 
(\ref{qcd1}) with (\ref{bose}), we can see that the latter flattens out at low values of $x_F$, because of the
mass $m_{J/\psi}$, forming a ``plateau''. A spectrum 
from mechanism iii) is a little more difficult to predict. It involves the convolution of two distributions of
the type (\ref{bose}) (for the two colliding massless partons) and therefore we  expect it to be  more steeply 
falling with $x_F$. As it will be seen, within certain approximations, it is given by:
\begin{center}
\beq
\simeq  \,\,\,\, \frac{1}{x_F}
\label{qgp1}
\eeq
\end{center} 
This behavior was found with the  CEM. It remains to be shown that it
is not an artifact of the model and that  it would also be found in the thermal 
version of the COM. 
Comparing  (\ref{qcd1}), (\ref{bose}) and (\ref{qgp1}) we see that, if 
QGP is formed {\it and} if the multiplicity of ``in-plasma-formed'' $J/\psi$'s 
is large enough, we may expect to see an enhancement (or even a peak)  at the  
origin of the $x_F$ distribution. 
This signal is interesting because it was 
not predicted by any other production mechanism. However  it relies on a very 
accurate knowledge of the pQCD, thermal and QGP abundancies, which is 
very difficult, as it will be dicussed in detail later.

At this point one might argue that after being produced 
by the usual pQCD mechanism, the $c\bar{c}$  bound state suffers interactions 
either with the plasma or with comovers in a hadron gas. These interactions
(called here Final State Interactions) will distort the initial $x_F$ 
distribution of the bound state.
 In Ref. \cite{huf} it was shown 
that FSI (either with a plasma or a hadron gas) will suppress the charmonium 
yield at $x_F=0$, giving origin to a dip in the central $x_F$ region.  

We are suggesting that a peak will arise at $x_F=0$ as a consequence of in-plasma production. 
This is a signal which will emerge from the background composed by initial pQCD production. 
If we  ignore the final state interactions we will calculate a background which already contains 
an enhancement of $J/\psi$ production in the low $x_F$ region, making it harder to detect the signal. 
In other words, if we  treat correctly the FSI this will lead to a background 
with a dip around $x_F=0$ instead of an enhancement. Therefore, at least for the sake of our argument, neglecting FSI 
is being conservative. This is the correct procedure in order to isolate the new effect, which we are looking for.
Nevertheless, at the end of this work we will  include FSI quantitatively and investigate their effect on the  
$J/\psi$ momentum distribution.

The text is organized in the following way: 
in section II we describe direct production; in sections III and IV we 
address production in the plasma and
final thermal production respectively; in section V we discuss final state 
interactions;  in section VI we present numerical 
results and in the last section we make some concluding remarks.

\section{Initial $J/\psi$ production in nucleus-nucleus collisions}

Nowadays perturbative QCD calculations of $c\bar{c}$ production can be found in textbooks. Nevertheless, 
we shall, in what follows, give some formulas to introduce the notation, to stress the role played by 
shadowing and to obtain expressions in terms of $x_F$, the variable of interest. One popular
approach to charm production is the  
color evaporation model  \cite{ramona,ramona2,cem}. 
We are aware of the discrepancies between 
different implementations of perturbative QCD in charm production \cite{ramona2}. In particular, there is another model, namely the color octet model 
\cite{bbl}, which gives a more sophisticated description of the hadronization process in terms of non-perturbative matrix elements. While the COM seems to 
give a very good description of high $p_T$ data ($p_T > 2 $ GeV), 
especially those from the Fermilab Tevatron, it is not so reliable at 
smaller $p_T$ \cite{cho} ($p_T < 2 $ GeV), where the bulk of charmonium production takes place and where we wish to make a comparison between  QCD and 
plasma production. For our purpose of studying the $x_F$ distributions, as 
can be inferred from the calculations presented in \cite{gou}, the COM prediction is compatible with a smooth behavior of the cross section
$ d \sigma / d x_F$ at the origin $x_F \simeq 0$, much like the CEM. 
Therefore, for 
simplicity, we shall use this latter model in what follows. 
Our results will certainly 
depend on the choice of approach. Inspite of the uncertainties, we will assume that, since we are interested in 
the low $x_F$ region, pQCD is enough and nonperturbative effects are small, in sharp contrast to what happens 
at large $x_F$ \cite{nos}.  

In the CEM, charmonium is defined kinematically 
as a $c\bar{c}$ state with mass below the  $D\bar{D}$ threshold. In leading order (LO) the cross 
section is computed  with the use of perturbative QCD for the diagrams of the elementary processes 
$q\bar{q} \rightarrow  c\bar{c}$ and $gg \rightarrow c\bar{c}$ convoluted with the parton densities 
in the projectile  and in the  target. Calling $x_F$ the fractional momentum of the 
produced pair (with respect to the momentum of a projectile nucleon in cm frame) and $\sqrt{s}$ the 
cm energy of a nucleon-nucleon collision, the cross section for production of a $c\bar{c}$  pair 
with mass $m$ is just given by:
\begin{eqnarray}
\frac{d\sigma^{pp \rightarrow c\bar{c}}}{dx_F dm^2}
&=&\int_{0}^{1} \, dx_1 \, dx_2 \, \delta (x_1 x_2 s - m^2) \, \delta (x_F - x_1 + x_2) 
\, H(x_1,x_2;m^2) \nonumber \\
&=&\frac{1}{s \, \sqrt{x_F^2 + 4m^2/s}} \,  H(x_{01},x_{02};m^2) \,\,\,\,\,\,\,; 
\,\,\,\,\,\,\, x_{01,02}= \frac {1}{2} \left(\pm x_F + \sqrt{x_F^2 + 4m^2/s} \right)
\label{cem1}
\end{eqnarray}
where $x_1$ and $x_2$ are the nucleon momentum fractions carried respectively  by partons in the 
projetile and target. The function  $H(x_1,x_2;m^2)$, which represents the convolution of 
the elementary cross sections and parton densities is given by:
\begin{eqnarray}
H(x_1,x_2;m^2)&=& f_g(x_1,m^2) \, f_g(x_2,m^2) \, \hat{\sigma}_{gg}(m^2) 
\nonumber \\
&+& \sum_{q=u,d,s} \, [f_q(x_1,m^2) \, f_{\bar{q}}(x_2,m^2) + 
f_{\bar{q}}(x_1,m^2) \, f_q(x_2,m^2)] \, \hat{\sigma}_{q\bar{q}}(m^2) 
\label{cem2}
\end{eqnarray}
with the parton densities $f_i(x,m^2)$ in the nucleon computed at the scale $m^2=x_1 x_2 s$.

The LO elementary cross sections in terms of the pair invariant mass ($m$)
are given by \cite{gluck}:
\begin{equation}
\hat{\sigma}_{gg}(m^2) = \frac {\pi \alpha_s^2 (m^2)} {3 m^2}  
\left\{ \left(1 + \frac{4 m_c^2}{m^2} + \frac{m_c^4}{m^4} \right) 
ln \left[\frac{1+\lambda}{1-\lambda} \right]-\frac{1}{4} 
\left(7+\frac{31m_c^2}{m^2}\right) \lambda \right\}
\label{siggg}
\end{equation}
\begin{equation}
\hat{\sigma}_{q\bar{q}}(m^2) = \frac {8 \pi \alpha_s^2 (m^2)} {27 m^2}
\left(1 + \frac{2 m_c^2}{m^2}\right) \lambda \,\,\,\,\,\,\,\,\,\,;
\,\,\,\,\,\,\,\,\,\, \lambda=\left[1-\frac{4 m_c^2}{m^2}\right]^{1/2}
\label{sigqq}
\end{equation}
where $m_c$ is the mass of the $c$ quark.

The production  cross section of the charmed state $i$ ($=J/\psi, \, \psi'$
or $\chi_{cJ}$)  ${\sigma_i}$, is then  finally obtained by integrating the free pair 
cross section $c\bar{c}$ over the invariant mass $m$ starting from the production
threshold $2 m_c$ up to open charm production threshold  
$2 m_D$: 
\begin{equation}
\frac{d\sigma^{pp\rightarrow J/\psi }}{dx_F} = F_{J/\psi} \, \int_{4m_c^2}^{4m_D^2} \, dm^2
\frac{d\sigma^{pp\rightarrow c\bar{c}}}{dx_F dm^2}
\label{cem3}
\end{equation}
where $F_{J/\psi}$ is the fraction of $\sigma^{c\bar{c}}$ which contains the corresponding 
$c\bar{c}$ resonance.

This model describes well  the experimentally measured  $x_F$ distribution of hidden charm both 
with LO and NLO cross sections, provided that  $F_i^{LO}$ is defined as $F_i^{NLO}$ multiplied by
a theoretical factor $\kappa$, which is equal to the ratio of the NLO and LO cross sections
($F_{J/\psi}^{NLO} \approx 2,54 \%$) \cite{gavai}.

In what follows we shall use the CEM to study perturbative $J/\psi$  production in 
nucleus-nucleus collisions at RHIC. As it is well known, in nuclear collisions and in processes 
involving small values of $x$,  shadowing plays an important role  
(see, for example,  \cite{greiner2}). In this case  expression (\ref{cem2}) is rewritten as:
\begin{eqnarray}
H_{AB}(x_1,x_2;m^2)&=& A\,B\,\biggr\{f_g(x_1,m^2)\,R_g^A(x_1,m^2) \, f_g(x_2,m^2)\,R_g^B(x_2,m^2) 
\hat{\sigma}_{gg}(m^2) \nonumber \\
&+& \sum_{q=u,d,s}  \biggr[f_q(x_1,m^2)\,R_q^A(x_1,m^2) \, f_{\bar{q}}(x_2,m^2)\,R_{\bar{q}}^B(x_2,m^2) \nonumber \\
&+& f_{\bar{q}}(x_1,m^2)\,R_{\bar{q}}^A(x_1,m^2) \, 
f_q(x_2,m^2)\,R_q^B(x_2,m^2)\biggr] \, 
\hat{\sigma}_{q\bar{q}}(m^2)\biggr\}
\label{cem4}
\end{eqnarray}
where
\beq
R_i^{A}(x,m^2)=\frac{f_{i}^{A}(x,m^2)}{f_{i}(x,m^2)}
\label{ra}
\eeq
with $f_{i}^{A}(x,m^2)$ being the $i$ parton momentum distribution in a nucleon inside the nucleus $A$. 
Replacing $H$ by $H_{AB}$ in (\ref{cem1}) we obtain the cross section for $c\bar{c}$ production in $A-B$ 
collisions:
\begin{eqnarray}
\frac{d\sigma^{AB \rightarrow c \bar{c}}}{dx_F} \,  &=& 
\, \kappa \, \int_{4m_c^2}^{\infty} \, dm^2\,
\frac{1}{s \, \sqrt{x_F^2 + 4m^2/s}} \,  H_{AB}(x_{01},x_{02};m^2)
\label{cemnumbercc}
\end{eqnarray}
and  the  $J/\psi$ production cross section as:
\begin{eqnarray}
\frac{d\sigma^{AB \rightarrow J/\psi}}{dx_F} \,  &=& 
\, \kappa \, F_{J/\psi}^{NLO} \, \int_{4m_c^2}^{4m_D^2} \, dm^2\,
\frac{1}{s \, \sqrt{x_F^2 + 4m^2/s}} \,  H_{AB}(x_{01},x_{02};m^2)
\label{cem5}
\end{eqnarray}

In a central $A+A$ collision the number of $J/\psi$'s  produced directly is related to the cross section  
by \cite{eskola3}:
\begin{equation}
\frac{d N_{direct}^{AA \rightarrow J/\psi}}{d x_F} \cong  \frac{1}{\pi\,R_A^2} \,  
\frac{d \sigma^{AA\rightarrow J/\psi}}{d x_F} 
\label{npsiqcd}
\end{equation} 
where $R_A$ is the radius of nucleus $A$. 

The expression above would give the final distribution if there were no absorption due to further interactions
with partons in a QGP or hadrons in a comoving fireball. The effect of these final state interactions will be 
discussed later.  

Since we are going to compare the total number of initially produced $ c \bar{c}$ pairs and 
$J/\psi$'s with those produced in the plasma it is useful to introduce the compact notation 
\begin{equation}
N^{c \bar{c}}_{QCD} = \frac{1}{\pi \, R_A^2} \, 
\int_{0}^{1} \, d x_F \, \frac{d\sigma^{AA \rightarrow c \bar{c}}}{dx_F}
\label{ccbarqcd}
\end{equation}
for the number of $ c \bar{c}$ pairs and 
\begin{equation}
N^{J/\psi}_{QCD} = \int_{0}^{1} \, d x_F \, \frac{d N_{direct}^{AA \rightarrow J/\psi}}{d x_F} 
\label{psiqcd}
\end{equation}
for the number of $J/\psi$'s. 

Expressions (\ref{ccbarqcd}) and (\ref{psiqcd}) are calculated with the nuclear parton distributions which 
take shadowing into account. While this makes calculations more realistic, it also introduces some model 
dependence in the results. In order to have a baseline for comparisons it is useful to introduce the 
equivalent definitions of $N^{c \bar{c}}_{QCD}$ and $N^{J/\psi}_{QCD}$ without shadowing. For central $A-A$ 
collisions  these are:
\begin{equation}
N^{c \bar{c}}_{QCD} \, =   T_{AA}(b=0) \, \int_{0}^{1} \, d x_F  \, 
\frac{d \sigma^{pp\rightarrow c \bar{c}}}{d x_F}    \,\,\,\, \mbox{ \,\,\,  (no shadowing)}
\label{ccbarnoshad}
\end{equation}
and
\begin{equation}
N^{J/\psi}_{QCD} \, =  \,  F_{J/\psi}^{NLO} \,  T_{AA}(b=0) \, \int_{0}^{1} \, d x_F  \, 
\frac{d \sigma^{pp\rightarrow c \bar{c}}}{d x_F}  \,\,\,\, \mbox{ \,\,\,  (no shadowing)}
\label{psinoshad}
\end{equation}
where $T_{AA}$ is the usual nuclear overlap function.
In Table I we present some quantitative results for (\ref{ccbarqcd}), (\ref{psiqcd}), 
(\ref{ccbarnoshad}) and (\ref{psinoshad}). As it can be seen, these numbers change with different  choices
for the charm quark mass and for  the parton distribution. Inclusion of shadowing reduces the number of 
charm quark pairs and $J/\psi$'s in about 10 \%,  choosing different parton densities may  change these 
numbers in 40 \% but  what changes the most our results is the adopted value for the charm quark mass. This
choice may alter the numbers by a factor 4.   Our number of $c \bar{c}$ pairs is smaller but still compatible 
with other estimates presented, for example, in \cite{ramona2}. On the other hand, the number of $J/\psi$'s 
can be  estimated  taking into account the recent PHENIX data \cite{phenix} on $p - p$ collisions. These estimates 
\cite{cassing} indicate that $N^{J/\psi}_{QCD} \simeq 0.1$, which is close to 
the number of $J/\psi$'s in Table I.

\section{$J/\psi$ production in the plasma}

\subsection{The number of $c\bar{c}$ pairs}

As in the previous section we reproduce below (to a great extent) textbook material \cite{wong}, with the 
purpose of defining our 
notation. The derivation presented here is also useful because, relaxing the screening hypothesis, we calculate
the production of charm {\it bound states} in the plasma and we integrate the rate on all variables except $x_F$. 
All this requires some straightforward but  not very often shown manipulations.

The computation of the in-plasma $c\bar{c}$ pair production rate goes back to the late eighties \cite{shor}, was 
discussed in short papers (for example \cite{wang}),   included in comprehensive review articles \cite{raf} 
during the nineties and experienced improvements due to advances in thermal field theory \cite{levai}.

Assuming that QGP is formed, we then have a gas of quarks and gluons with momenta obeying  respectively 
Fermi-Dirac and Bose-Einstein distributions that can collide producing $c\bar{c}$ pairs. This is the 
same mechanism, which in the strange sector causes the strangeness enhancement. In the charm case, there 
will be always a ``charm enhancement'' but, because of screening, the charmed quarks will mostly form 
open charm.  In the present calculation we will let part of the thermally produced charm Coulomb bound states 
escape. Since the temperature is high, there will be a significant number of parton-parton collisions 
at a cms energy high enough to produce charm quark pairs, which may form charmonium states.  We will now estimate 
their production rate using the CEM in a thermal environment.

The charmonium production rate in the  reaction $g g \rightarrow c\bar{c}$, at temperature $T$ 
is given by \cite{wong}:
\begin{equation}
\frac {dN^{g g \rightarrow c \bar{c}}}{dt d^3 x} =  \frac{1}{2} \, \frac{1 }{(2 \pi)^6} \, 
g^2_{g} \, \int d^3p_1 d^3p_2 \, f_g (E_1) \, f_g (E_2) \, \hat{\sigma}_{g g}^{LO}
(m^2) \, v_{12}
\label{dndtd3xgg}
\end{equation}
where $g_g$ is the gluon statistical factor (number of colors $\times$ number polarization states)
$v_{12}$  is the relative velocity between colliding gluons with energies 
$E_1$ and $E_2$ and three momenta $\vec{p_1}$ and $\vec{p_2}$,  $\hat{\sigma}_{g g}^{LO}$ is the 
elementary gluon-gluon cross section  (\ref{siggg}) and  $f_g (E_i)$ the usual thermal distribution 
function:
\begin{equation}
f_g (E_i) = \frac {1}{e^{E_i/T}-1}
\label{fgde}
\end{equation}

We now introduce the charm pair four momentum $p$ with help of the delta function  $\delta ^4 [p-(p_1+p_2)]$: 
\begin{equation}
\frac {dN^{g g \rightarrow c \bar{c}}}{d^4 p}  =  \frac{1}{2} \, \frac{1 }{(2 \pi)^6} \,  g^2_{g} \, 
\int dt d^3 x d^3p_1 d^3p_2 \, f_g (E_1) \, f_g (E_2) \, \hat{\sigma}_{g g}^{LO}
(m^2) \, v_{12} \,\delta ^4 [p-(p_1+p_2)]
\label{dnd4pgg}
\end{equation}

In the expression above we have:
\begin{equation}
p_1 \equiv (E_1,\vec {p}_{T_1},p_{z_1})\,\,\,\,;\,\,\,\,
p_2 \equiv (E_2,\vec {p}_{T_2},p_{z_2})\,\,\,\,;
\,\,\,\,p \equiv (E,\vec {p}_{T},p_{z})
\label{quadrips}
\end{equation}
\begin{equation}
 m^2 =(p_1+p_2)^2 \,\,\,\,;\,\,\,\,v_{12}=\frac{1}{2} \frac{m^2}{E_1 E_2}
\label{m2v12}
\end{equation}
where $m$ is the invariant mass of the pair. We next decompose the delta function into temporal, transverse and
longitudinal components:
\begin{eqnarray}
\frac {dN^{g g \rightarrow c \bar{c}}}{dm^2 d\vec{p}_T dp_z}  &=&  \frac{1 }{4(2 \pi)^6} \,  g^2_{g} \, 
\int dt d^3 x \frac{1}{E_1}\,d^3p_1 \frac{1}{E_2}\,d^3p_2 \, f_g (E_1) \, f_g (E_2) \, \hat{\sigma}_{g g}^{LO}(m^2) 
\, m^2 \nonumber \\
&\times& \,\delta [m^2-(p_1+p_2)^2]\,\delta ^2 [\vec {p}_{T}-(\vec {p}_{T_1}+\vec {p}_{T_2})]\,
\delta [p_z-(p_{z_1}+p_{z_2})]
\label{dndm2ptpzgg1}
\end{eqnarray}

Making use of the identity
\begin{equation}
\int \frac{1}{2 E_i}\,d^3p_i = \int d^4p_i \delta(p_i^2) \theta (E_i) =
\int dE_i d\vec {p}_{T_i} dp_{z_i}
\delta [E_i^2-p_{t_i}^2-p_{z_i}^2] \theta (E_i) 
\end{equation}
we arrive at:
\begin{eqnarray}
\frac {dN^{g g \rightarrow c \bar{c}}}{dm^2 d\vec{p}_T dp_z}  &=&  \frac{1}{(2 \pi)^6} \,  g^2_{g} \,
\int dt d^3 x \int_0^\infty dE_1 \int_{-\infty}^\infty d\vec {p}_{T_1} 
\int_{-\infty}^\infty dp_{z_1}\,\delta [E_1^2-p_{t_1}^2-p_{z_1}^2]\,\theta (E_1)  \nonumber \\
&\times&\int_0^\infty dE_2 \int_{-\infty}^\infty d\vec {p}_{T_2} 
\int_{-\infty}^\infty dp_{z_2}\,\delta [E_2^2-p_{t_2}^2-p_{z_2}^2]\,\theta (E_2)
\, f_g (E_1) \, f_g (E_2) \nonumber \\
&\times&\,\, \hat{\sigma}_{g g}^{LO}(m^2) \, m^2\,\delta [m^2-(p_1+p_2)^2]\,\delta ^2 [\vec {p}_{T}
-(\vec {p}_{T_1}+\vec {p}_{T_2})]\,\delta [p_z-(p_{z_1}+p_{z_2})]
\label{dndm2ptpzgg2}
\end{eqnarray}

Integrating  in $p_{z_2}$, $p_{T_1}$, $p_{T_2}$ and $E_2$ and defining $\beta$ as the 
angle between $\vec {p}_{T}$ and $\vec {p}_{T_1}$, we finally obtain:
\begin{eqnarray}
\frac {dN^{g g \rightarrow c \bar{c}}}{dp_z} &=& \frac{\pi}{4(2 \pi)^6} \, g^2_{g} \,
\int _{4m_c^2}^{\infty} dm^2 \int_0^\infty dp_T^2 \int_0^{2\pi} d\beta \int dt d^3 x
\int_0^\infty dE_1 \int_{-\infty}^\infty dp_{z_1} f_g (E_1) \, f_g (E-E_1) \nonumber \\
&\times& \hat{\sigma}_{g g}^{LO}(m^2)\,\frac {m^2}{E} 
\,\delta [m^2-2EE_1+2p_T(E_1^2-p_{z_1}^2)^{1/2} \cos(\beta)+2p_zp_{z_1}]\nonumber \\
&\times& \theta (E_1)\, \theta (E-E_1)\, \theta (E_1^2-p_{z_1}^2)\nonumber \\
&=& \frac{\pi}{4(2 \pi)^6} \, g^2_{g} \,
\int _{4m_c^2}^{\infty} dm^2 \int_0^\infty dp_T^2 \int_0^{2\pi} d\beta \int dt d^3 x
\int_0^\infty dE_1 f_g (E_1) \, f_g (E-E_1) \nonumber \\
&\times&\,\hat{\sigma}_{g g}^{LO}(m^2)\, \frac {m^2}{E} \,
\left [ \frac {1}{|H(h_1)|} \theta (E_1^2-h_1^2) + \frac {1}{|H(h_2)|} 
\theta (E_1^2-h_2^2) \right ] \,\theta (E_1) \theta (E-E_1)
\label{dndpzgg}
\end{eqnarray}
where $E$, $H(h_{1,2})$ and  $h_{1,2}$ are given by:
\begin{eqnarray}
E=[m^2+p_T^2+p_z^2]^{1/2}\,\,\,\,;\,\,\,\,
H(h_{1,2})=2p_z-\frac {2p_T\cos(\beta)}{(E_1^2-h_{1,2}^2)^{1/2}} h_{1,2} 
\label{eeagazao}
\end{eqnarray}
\begin{eqnarray}
h_{1,2}&=&\frac{1}{2} \frac{1}{p_z^2+p_T^2 \cos ^2(\beta)} \biggr\{p_z(2EE_1-m^2) 
\mp  \{p_T^2 \cos ^2(\beta) \nonumber \\
&\times& [4E_1^2(p_T^2 \cos ^2(\beta)+p_z^2)-(m^2-2EE_1)^2]\}^{1/2}\bigg\}
\label{h12}
\end{eqnarray}
In the first line of (\ref{dndpzgg}) we have $\delta [m^2-2EE_1+2p_T(E_1^2-p_{z_1}^2)^{1/2} \cos(\beta)+2p_zp_{z_1}]$. 
It is easy to see that  taking either $E_1 \simeq p_{z_1}$ or  $E_1>>p_{z_1}$  and using the $\delta$ function to perform 
the integral in $p_{z_1}$ we obtain a factor $1/p_z$. This factor survives and, at the end, gives the $1/x_F$ 
structure (\ref{qgp1}) mentioned in the introduction. The actual numerical calculation involves no approximation and
has a similar behavior.  Notice that, because of the kinematical cuts in the  invariant mass integral, we are now 
restricting the result to the rate to charmonium states. 
A similar expression can be derived for $c\bar{c}$ production originating from 
quark-antiquark annihilation:
\begin{eqnarray}
\frac {dN^{q\bar q \rightarrow c \bar{c}}}{dp_z} &=& \frac{\pi}{2(2 \pi)^6} \,
 g^2_{q ({\bar q})} \,\sum_{q=u,d,s} \int _{4m_c^2}^{\infty} dm^2 \int_0^\infty dp_T^2 \int_0^{2\pi} d\beta 
\int dt d^3 x \int_0^\infty dE_1 f_q (E_1) \, f_{\bar q} 
(E-E_1) \nonumber \\
&\times&\,\hat{\sigma}_{q {\bar q}}^{LO}(m^2)\, \frac {m^2}{E} \,
\left [ \frac {1}{|H(h_1)|} \theta (E_1^2-h_1^2) + \frac {1}{|H(h_2)|} 
\theta (E_1^2-h_2^2) \right ] \,\theta (E_1) \theta (E-E_1)
\label{dndpzqqbar}
\end{eqnarray}
where $g_{q(\bar{q})}$ is the statistical factor for quarks (antiquarks) (number of colors $\times$ 
number of spin states),  $\hat{\sigma}_{q {\bar q}}^{LO}(m^2)$ is given by   (\ref{sigqq}) and 
$f_{q,\bar q} (E_i)$ is the usual quark (antiquark) thermal distribution function:
\begin{equation}
f_{q,\bar q} (E_i) = \frac {1}{e^{E_i/T}+1}
\label{fqqbarde}
\end{equation}

In order to account for  expansion effects  we shall assume that the system cools down following  
Bjorken hydrodynamics, in which  temperature and proper time are related by \cite{wong}:
\begin{equation}
\frac{T(\tau)}{T_0}=\left(\frac{\tau_0}{\tau}\right)^{1/3}
\end{equation}
where $T_0$ is the initial temperature and $\tau_0$ is the thermalization time, which marks the beginning of the
hydrodynamical expansion. With the help of this relation we can 
perform the following change of variables:
\begin{equation}
dt d^3 x = dx_\perp \, \tau \, dy \,d\tau = -3\, dV \,\tau_0\,T_0^3\,T^{-4}\,dT
\,\,\,\Longrightarrow\,\,\,
\int dt d^3 x = 3\,\int_{T_f}^{T_0} \tau_0\,T_0^3\,T^{-4}\,V(T)\,dT
\end{equation}

Finally we introduce the energy scale $\sqrt s$ (the invariant energy of a single nucleon-nucleon collision): 
\begin{equation}
p_z=x_F \frac {\sqrt s}{2}
\label{pzxf}
\end{equation}

Expressions (\ref{dndpzgg}) and (\ref{dndpzqqbar}) can thus be rewritten as:
\begin{eqnarray}
\frac {dN^{g g \rightarrow c \bar{c}}}{dx_F} &=& \frac{3\pi}{8(2 \pi)^6} \, g^2_{g} \,
\sqrt s \, \kappa \,\int _{4m_c^2}^{\infty} dm^2 \int_0^\infty dp_T^2 \int_0^{2\pi} d\beta 
\int_{T_f}^{T_0} \tau_0\,T_0^3\,T^{-4}\,V(T)\,dT
\int_0^\infty dE_1  \nonumber \\
&\times&\,f_g (E_1) \, f_g (E-E_1)\,\hat{\sigma}_{g g}^{LO}(m^2)\, \frac {m^2}{E} \,
\left [ \frac {1}{|H(h_1)|} \theta (E_1^2-h_1^2) + \frac {1}{|H(h_2)|} \,\theta (E_1^2-h_2^2) \right ]
 \nonumber \\
&\times& \,\theta (E_1) \theta (E-E_1)
\label{dndxfgg}
\end{eqnarray}
\begin{eqnarray}
\frac {dN^{q\bar q \rightarrow c \bar{c}}}{dx_F} &=& \frac{3\pi}{4(2 \pi)^6} \,
 g^2_{q ({\bar q})} \,\sqrt s \, \kappa \, \sum_{q=u,d,s} \int_{4m_c^2}^{\infty} dm^2 
\int_0^\infty dp_T^2 \int_0^{2\pi} d\beta 
\int_{T_f}^{T_0} \tau_0\,T_0^3\,T^{-4}\,V(T)\,dT
\int_0^\infty dE_1  \nonumber \\
&\times&\,f_q (E_1) \, f_{\bar q} (E-E_1)\,\hat{\sigma}_{q {\bar q}}^{LO}(m^2)\, \frac {m^2}{E} \,
\left [ \frac {1}{|H(h_1)|} \theta (E_1^2-h_1^2) + \frac {1}{|H(h_2)|} 
\theta (E_1^2-h_2^2) \right ] \nonumber \\
&\times&\,\theta (E_1) \theta (E-E_1)
\label{dndxfqqbar}
\end{eqnarray}

The volume of the system evolves in time according to:
\begin{equation}
V(\tau)=V_0 \frac{\tau}{\tau_0}\,\,\,\,\,\,\,\Longrightarrow\,\,\,\,\,\,\,
V(T)=V_0\left(\frac{T_0}{T}\right)^3
\end{equation}
where $V_0=\pi\,R_A^2\,\tau_0$,  $R_A = (1.18\,A^{1/3}-0.45)$ fm and the $\kappa$ factor was introduced 
explicitly.

Summing Eqs. (\ref{dndxfgg}) and (\ref{dndxfqqbar}) and integrating over $x_F$ we obtain the total number of 
in-plasma produced  $c \bar{c}$ pairs, which we call $N^{c \bar{c}}_{QGP}$.

\subsection{Comparison with other works}

We shall now compare the number of   $c \overline{c}$ obtained in this work   with the number
of charmed pairs obtained in other works. There are some known  papers on  the subject 
\cite{wang,levai,lmw,raf,str}. 
From the reading of these papers, we conclude that there are large 
discrepancies in the numbers and in the way to obtain them. The sources of these 
discrepancies are:

\noindent
a) initial temperature of the plasma, $T_0$ (ranging from $300$ MeV to $550$ MeV) \\
b) degree of parton equilibration (described by the fugacity factors) \\
c) initial volume and/or thermalization time, $V_0$ and $\tau_0$ \\
d) total energy contained in the fireball \\
e) use or not of a $\kappa$ factor (=2) in computing thermal rates \\
f) use or not of temperature dependent $\alpha_s$ \\
g) mass of the charm quark $m_c$ (going from $1.2$ to $1.5$ GeV) \\
h) use or not of thermal masses for gluons and quarks in the reactions 
$g + g  \rightarrow c + \overline{c}$ and $ q + \overline{q} \rightarrow c + \overline{c}$. \\

Depending on the choices that one has to make in dealing with a) $\rightarrow$ h) 
the final value of in-plasma produced $c\bar{c}$ pairs can change by  orders 
of magnitude, going roughly  from  $N^{c \bar{c}}_{QGP} = 0.02$ in \cite{wang} to up $15$ 
in \cite{raf}.  In 
a comprehensive analysis this same variation was found in \cite{lmw}. In \cite{levai}, the 
authors arrive to the conclusion that in-plasma production was only a factor 
two smaller than the initial production. However in that paper the factor $\kappa = 2$ 
was not included in calculation. If it were, then $N^{c \bar{c}}_{QGP} \simeq N^{c \bar{c}}_{QCD}$.

 The list of uncertainties given above is not
unique and could be enlarged. Each of the items implies taking a decision or 
making an assumption. Since the list is already large it seems that a high  
precision calculation is hopeless. In fact the situation is not so bad since 
our knowledge will increase both because of new experimental results and 
because of lattice results.  RHIC data will impose severe constraints on 
items from a) to d). For example, taking together
data from all the four collaborations, we may expect to know with sufficient 
accuracy the rapidity distribution of charged particles, from which we can 
have a good knowledge of the total energy contained in the fireball 
(item d)). The global analysis of data on rapidity, $p_T$ spectra, 
abundancies, eliptic flow and HBT interferometry, will eventually rule out several 
initial conditions used in hydrodynamical  models and we will have much less 
uncertainty in the initial temperature of the plasma. On the theoretical side,
the perturbative QCD analysis of other processes
like, for example, $J/\psi$ photoproduction or $J/\psi$ production in 
$e^- e^+$ collisions may significantly reduce the uncertainty in the charm 
quark mass (item g)). Lattice calculations at finite temperature will 
hopefully reduce the freedom in the choices of items f) and h). A review of
what we may learn in the field in a near future can be found in the summary 
talk presented by Pisarski in the last Quark Matter meeting \cite{rob}. 

We have explicitly  investigated the effect of changing the mass of the quark $c$ and the differences which 
arise when we use the coupling constant running with $m^2$:
\beq
\alpha_s(m^2)=\frac{12 \pi}{(33 - 2 N_F) \ln{\frac{m^2}{\Lambda_{QCD}^2}}}
\label{alfa_m}
\eeq
or with $T$ \cite{levai}:
\beq
\alpha_s(T)=\frac{6 \pi}{(33 - 2 N_F) \ln{\frac{19 T}{\Lambda_{\overline{MS}}}}}
\label{alfa_T}
\eeq
where $N_F=3$, $\Lambda_{QCD}=230$ MeV and $\Lambda_{\overline{MS}} = 80$ MeV.

In Table II we present our results for the number of ``in-plasma born''  $c\bar{c}$ pairs for different 
values of couplings and charm quark masses. All the calculations were done with an initial plasma temperature of 
$T_0 =550$ MeV, $\alpha_s(M^2)$  and  $\alpha_s(T)$ are given by (\ref{alfa_m}) and (\ref{alfa_T}) respectively. 
The numbers inside parenthesis correspond to the  choice $\kappa =1$. Otherwise the numbers were obtained with 
$\kappa = 2$.

The fourth line of Table II shows that our results may change by a 
factor $30$ depending on the inputs used. Since there is no strongly preferred value for the $c$ quark mass, 
neither for $\kappa$ or for the functional form of  $\alpha_s$, we are not able to a priori discard any of 
these choices and our final results will reflect these uncertainties.  In some cases a direct comparison  with 
other works is possible. For example,  in the second column, comparing the numbers in parenthesis, we notice that 
we obtain nearly five times more pairs than in \cite{levai}. A similar excess is observed comparing the numbers in
parenthesis in the fourth column. Comparing our work with Ref. \cite{levai} we can see that the pair production 
mechanism is quite similar but the treatment given to the plasma expansion is different. Whereas we have used  the 
standard Bjorken  hydodynamics, in Ref. \cite{levai} a new hydodynamical model was introduced. Comparing the details 
of both approaches we concluded that in Ref. \cite{levai} the expansion and cooling of the system is much faster 
than in Bjorken hydrodynamics. Consequently the system stops much earlier to create $c \bar{c}$ pairs and the final
yield will be smaller.  We think that the expansion of the plasma must be included in any serious calculation, but
hydrodynamics is in itself a very complex subject. A prudent strategy, which we adopt here, is to study  the desired 
effect (charm production) first with the most standard and simple hydrodynamical model  and then, as a second step, 
plugg the charm production formalism into a state-of the-art hydrodynamical code. We will leave this last step for 
the future.  The comparison of the other results in Table II with other works shows that for similar inputs we 
obtain numbers which are compatible  with those presented in Refs. \cite{lmw} and \cite{raf}.
 
\subsection{The number of $J/\psi$'s}

The number of $J/\psi$'s produced in the plasma can be obtained from Eqs. (\ref{dndxfgg}) and (\ref{dndxfqqbar}).
We must change the upper limit of integration in $m^2$ introducing a kinematical cut-off, i.e., making the 
replacement $\infty \rightarrow 4 m^2_D$. In doing so, we rename the superscripts in 
Eqs. (\ref{dndxfgg}) and (\ref{dndxfqqbar}) to $g g \rightarrow J/\psi$ and  $ q \bar q \rightarrow J/\psi$. 
Moreover we introduce the  CEM multiplicative factor   $F_{J/\psi}$ and arrive at:
\begin{equation}
\frac {dN_{QGP}^{A A \rightarrow J/\psi}}{dx_F} =  f_s  \,   F_{J/\psi} \, 
[ \frac {dN^{g g \rightarrow J/\psi}}{dx_F} \, + \, 
\, \frac {dN^{q \bar q \rightarrow J/\psi}}{dx_F} ] 
\label{npsiqgp}
\end{equation}
and
\begin{equation}
N_{QGP}^{J/\psi} = \int^1_0 \,  d x_F \, \frac{dN_{QGP}^{A A \rightarrow J/\psi}}{dx_F}
\label{psiqgp}
\end{equation}

The color evaporation model  has an "intrinsic efficiency" given by the 
fractional factor $F_{J/\psi}$ above, which was fixed \cite{ramona2,ramona} in the analysis of $p-p$ reactions
to be  $F_{J/\psi} \simeq 0.02 = 2\%$. It is by no means obvious that the same value should hold for $A - A$ 
collisions. For simplicity, we shall assume that $F_{J/\psi}$ is universal and holds even for in-plasma 
production.
The ``screening factor''  $f_s$ is thus the only free number introduced here. It gives the probability that 
a $J/\psi$ formed inside the plasma survives the passage through the medium. In other words, $f_s$ 
accounts for dynamical screening, in which gluons in the plasma destroy the charmonium bound state. It varies from
$0$ to $1$.  A literal interpretation of \cite{msatz} or \cite{karsch} would imply $f_s=0$. We will however tolerate 
a small value of $f_s$ and examine the consequences. In fact, it is impossible to say what a "realistic" 
value of $f_s$ would be. Our choices,  based on some  numerical estimates,  are in the following interval:
\begin{equation}
10^{-3} \, \leq   \, f_s \, \leq  \, 10^{-2}
\label{intervalfs}
\end{equation}

If $f_s < 10^{-3}$ we  do not observe any visible effect of the in-plasma production. On the other hand, 
$f_s = 10^{-2}$,  which can be interpreted as meaning that  
1 \% of the in-plasma born $J/\psi$'s can survive as bound states, can be taken 
as an upper limit, beyond which, rather than taking small fluctuations into 
account we would be really challenging the well established concept of screening.

\section{ Final thermal $J/\psi$ production}

We will also, for completeness, consider the case where the $J/\psi$'s produced in the early stage of the 
collision  traverse the plasma being destroyed and then ``regenerated'' \cite{rapp}.  We call them ``thermal''. 
As has been discussed in the literature \cite{munz,greiner1}, regenerated $J/\psi$'s follow a thermal distribution 
with the 
temperature of the quark-hadron phase transition $ T_c \simeq 170$ MeV:
\beqa
\frac {dN^{A A \rightarrow J/\psi}_{thermal}}{dx_F} &=&  
\frac{\pi \sqrt{s} V_0 A}{2} \int_0^s d p_T^2 \,\, exp\,\,\left[-\frac{1}{T_c} \,(x_F^2\,s/4 +p_T^2 +
m_{J/\psi}^2)^{1/2}\right]
\nonumber \\*[7.2pt]
&=&\,\, const \, \, exp\, \left[ -\frac{1}{T_c} \,\left( \sqrt{x_F^2\,s/4 + m_{J/\psi}^2}\, \right) \right] \,\,
\left[1 + \frac{1}{T_c} \, \sqrt{x_F^2\,s/4 + m_{J/\psi}^2}\, \right]
\label{dnthermal}
\eeqa
where the constant in front of the integral  will be fixed later through the  normalization condition:
\begin{equation}
N_{th}^{J/\psi} = \int^1_0 \,  d x_F \, \frac{dN_{thermal}^{A A \rightarrow J/\psi}}{dx_F}
\label{psiterm}
\end{equation}

\section{Final state interactions}

In sections II, III and IV we discussed $J/\psi$ production. 
After being produced
these states will suffer interaction with the free partons of the plasma and
later with the hadronic comovers. In \cite{huf}, the final state interactions
(FSI) were incorporated (to the rate analogous to our  
(\ref{npsiqcd})) by the introduction of the multiplicative suppression factors
$S^{COM}_{FSI}$ and $S^{QGP}_{FSI}$, for interactions with comovers and 
QGP partons respectively. We shall adopt here the same procedure. In fact, 
in the case
of the initial QCD production, since we are using the same formalism, for the sake 
of comparison we shall later borrow the final corrected expression from 
\cite{huf}.  In the case of (\ref{psiqgp}), the $J/\psi$ dissociation by 
partons within the plasma is already taken into account by the ``screening 
factor'' $f_s$.  If the charmonium state survives the plasma phase, it will 
interact with hadronic comovers and (\ref{psiqgp}) will be multiplied by the
suppression factor \cite{cap,huf}:
\begin{equation}
S^{COM}_{FSI}  =  exp \left[ -\sigma_{co} n_{co} 
\ln \left( \frac{n_{co}}{n_{fo}} \right) \right]
\label{fsi}
\end{equation}
where $n_{co}$ is the comover density, $n_{fo}$ is the freeze-out density 
($=1.15 fm^{-2}$) and $\sigma_{co} \simeq 1 - 4$ mb 
is the charmonium-hadron cross section \cite{nos1,nos2}.  The above expression 
depends on the impact parameter, on the collision energy and on time at which 
the charmonium interacts with a given  comover. Moreover, since $n_{co}$ is a 
function of the rapidity, it will depend also on $x_F$. However, 
restricting the analysis to central collisions and to a fixed ($\sqrt{s}$) 
energy, the suppression factor tends to be constant, as concluded in 
\cite{rapp,cap,poll}. For simplicity we shall take here the average of the 
values quoted in these works:
\begin{equation} 
S^{COM}_{FSI}  = 0.25 
\label{smean}
\end{equation}
Final state interactions with comovers will be included in the thermal 
production differential rate (\ref{dnthermal}) in  the same way and with 
the use of (\ref{smean}).

\section{Feynman momentum distributions}

In Fig. 1 we show, for $Au-Au \, (\sqrt s = 200$ GeV) (1a) and $Pb-Pb \, (\sqrt s = 17$ GeV) (1b) 
collisions, the results obtained with (\ref{npsiqcd}) (dash-dotted line), (\ref{npsiqgp}) (solid line) 
and (\ref{dnthermal}) (dashed line).  
For the QCD distribution, we have used the nuclear parton distribution functions (PDF) 
parametrized as in \cite{eskola1}. In doing so,  the proton PDF's were taken from  \cite{GRV98} (GRV98 LO). 
We also used $m_c=1.2$ GeV, $m_D=1.87$ GeV and $\kappa=2$. With these choices the total number of produced 
$J/\psi$'s is determined. In computing (\ref{npsiqgp}) we have used $\tau_0=0.7$ fm  and  $T_0=550$ MeV, for $Au+Au$ and   
$\tau_0=1$ fm and $T_0=300$ MeV for $Pb + Pb$ collisions. $T_c$ was taken to be $170$ MeV.

In each figure all curves are normalized to the same number of produced $J/\psi$'s, which is given by (\ref{psiqcd}). 
In Fig. 1 we have used the  
parton distributions of \cite{eskola1}, obtaining   $N_{J/\psi} \simeq 0.09$ at RHIC and $N_{J/\psi} \simeq 0.001$ 
at SPS. 

Although the normalization is still artificial (it overestimates the 
number of $J/\psi$'s produced in QGP) it anticipates our main claim: QGP production will create a peak at 
$x_F \simeq 0$.  Even if suppressed, it will leave a signal.  Moreover, this may happen at RHIC and at SPS 
as well.

In order to further investigate the differences between the three $J/\psi$ production mechanisms
we  will study the relation between the measured number of $J/\psi$'s and the naive expectation 
based on what we know from $p p$ collisions. Experimentally this ratio is easily constructed from the measured 
$x_F$ spectra in $A + A$  and $p + p$  collsions. Theoretically, the numerator can be written in terms of our 
theoretical prejudices. 

As a starting point, we assume  that all $J/\psi$'s are produced directly and we  show  in Fig. 2 the following ratio: 
\begin{equation}
R_{I}(x_F) = \frac{\frac{1}{\pi R^2_A} \frac{d \sigma^{AA\rightarrow J/\psi }_{QCD}}{d x_F}}
    {T_{AA}(b=0) \frac{d \sigma^{pp\rightarrow J/\psi }}{d x_F}} =
\frac{\frac{d N_{direct}^{AB \rightarrow J/\psi}}{d x_F}}{T_{AA}(b=0) 
\frac{d \sigma^{pp\rightarrow J/\psi }}{d x_F}}
\label{r1}
\end{equation}
where the denominator is given by the product of (\ref{cem3}) with the nuclear overlap function $T_{AA}(b=0)$  
$\, = 29.9$ mb$^{-1}$ at RHIC and  $\, = 30.4$ mb$^{-1}$ at SPS.  
The numerators  were obtained with (\ref{cem5}), (\ref{npsiqcd})  and the nuclear parton distribution functions (PDF) 
parametrized as in \cite{eskola1} (solid line) and as in \cite{kumano} (dashed line). In the first case the 
proton PDF were taken from  \cite{GRV98} (GRV98 LO) whereas in the second case they were taken from \cite{MRST} 
(MRST LO). With this last set of parton distributions  $N_{J/\psi} \simeq 0.06$ at RHIC and $N_{J/\psi} \simeq 0.001$ 
at SPS.

$R_{I}$ shows essentially the effect of shadowing in the low $x$ region. We can see that if no effect is 
present, other than shadowing, $J/\psi$ spectrum  produced in  $A A$ collisions is just a constant  times the
corresponding $p p$ spectrum. As it can be seen this constant may change a lot with the PDF.  
This constancy in the region $x_F < 0.1$  will be important for the subsequent discussion. This figure shows also 
that the value of the constant decreases when we go from SPS to RHIC collisions. This same behavior was found in 
\cite{huf}.

In Figure 3 we present the  $x_F$ distribution of the in-plasma produced $J/\psi$'s. The screening factor
will be fixed to $f_s = 10^{-2}$.  The curves are obtained with expression (\ref{npsiqgp}) normalized to the
largest (solid line) and to the smallest (dashed line) number of $J/\psi$'s. These normalization constants 
correspond to the choices 2.7 and 0.16, in the last line of Table II. Fig. 3a and 3b correspond to RHIC and 
SPS collisions respectively.

If we believe that all production is mainly direct plus some small plasma component, then  (\ref{r1}) aquires 
the form: 
\begin{equation}
R_{II}(x_F) = \frac{ 
\frac{d N^{AA \rightarrow J/\psi }_{direct}}{d x_F} \, + \,
 \frac{d N^{AA \rightarrow J/\psi }_{QGP}}{d x_F}} 
{  T_{AA}(b=0)  \frac{d \sigma^{pp \rightarrow J/\psi }}{d x_F}}
\label{r2}
\end{equation}
where, for the moment, we neglect the final state interactions discussed in 
the previous section.
If the screening in the plasma is so strong that $f_s=0$, then, 
$R_{II} = R_{I} \simeq  const$. If, however, there is a small $J/\psi$  survival probability, for example 
$f_s = 10^{-2}$, there 
will be a noticeable change in $R_{II}$. This is illustrated in Fig. 4, where we  see a pronounced deviation 
from a flat behavior in the low $x_F$ region if there is a contribution from the plasma.  This interesting 
feature is due to nature of plasma production, which is peaked at $x_F \simeq 0$.  In Fig. 4 the two upper lines 
correspond to the largest plasma  $J/\psi$ production ($N^{J/\psi}_{QGP} = 2.7$) and the lower two lines to the 
smallest plasma contribution ($N^{J/\psi}_{QGP} = 0.16$). The scrrening factor is kept fixed to $f_s = 10^{-2}$. 
In all cases solid and dashed lines mean the same PDF's as in Fig. 2. 

Figs. 4a and 4b refer to $Au+Au$ at RHIC and  to $Pb+Pb$ at SPS respectively. 
We can clearly see that deviations from unity happen more strongly for RHIC collisions.

The above situation  has to be regarded as an extreme case, in which $J/\psi$'s are produced via parton fusion (with 
shadowing taken into account) in the initial state of the nucleus-nucleus collision and then just escape from the 
fireball, without further nuclear or comover suppression.

We now consider the opposite extreme case, where all 
$J/\psi$'s are formed in the final state. Final state here 
means that the original $J/\psi$'s are totally destroyed by plasma screening and recriated at the final stage of
the fireball life (at $T=T_c \simeq 170$ MeV) by coalescence and then just ``emerge ready'' during the fase transition,
obeying the thermal distribution (\ref{dnthermal}).  This scenario is supported by
the calculations performed in \cite{rapp,munz,abrs}. Assuming this production mechanism the ratio defined above is 
modified to:
\begin{equation}
R_{III}(x_F) = \frac{ S^{COM}_{FSI} 
 \frac{ d N^{AA \rightarrow J/\psi }_{Thermal}}{d x_F} \, 
+ \, S^{COM}_{FSI}    \frac{ d N^{AA \rightarrow J/\psi }_{QGP}}{d x_F}} 
{  T_{AA}(b=0)  \frac{d \sigma^{pp \rightarrow J/\psi }}{d x_F}}
\label{r3}
\end{equation}
where $ S^{COM}_{FSI} = 0.25 $. 
When discussing thermal production a crucial aspect is how to normalize the distribution (\ref{dnthermal}). The 
most straightforward procedure would be to normalize it in  such a way that the total number of $c$ and $\overline{c}$ 
quarks in the final state would match the number of charm quarks pairs produced initially and computed with the 
help of perturbative QCD. However, as we can see from Table I and also from the recent and comprehensive analysis  
performed  in Ref. \cite{ramona2}, 
already at the $p + p \rightarrow c \overline{c} + X$ level there is some uncertainty in the cross section 
coming from choices of the quark masses, renormalization scales and parton density parametrizations. A further  
source of uncertainty arises when we go from $p p $ to $A A$ collisions and introduce shadowing effects. Moreover, 
in a very recent calculation of thermal charm production \cite{abrs}, already taking into account the first PHENIX 
data \cite{phenix}, it was shown that, to explain the overall magnitude of the data, one needs to increase the NLO QCD  
$c \overline{c}$ yield by a factor $2.8$. In view of this lack of precise knowledge, in our study we will normalize  
the integral of Eq. (\ref{dnthermal}), $N^{th}_{J/\psi}$ in two different ways:
\begin{equation}
I) \,\, N^{J/\psi}_{th} = 3 \, N^{J/\psi}_{QCD} \,\,\,\,;\,\,\,\,
II)\,\, N^{J/\psi}_{th} = 0.5 \, N^{J/\psi}_{QCD}
\label{normal}
\end{equation}
where $ N^{J/\psi}_{QCD}$ is given by Eq. (\ref{psiqcd}), which, in turn, will be evaluated with two 
different nuclear parton distributions

In Fig. 5 we plot the ratio $R_{III}$. The two lower (upper) pannels refer to $Pb + Pb$ ($Au + Au$) data measured at 
SPS (RHIC). In the two left (right) pannels we use the Eskola {\it et al.} \cite{eskola1}  (Hirai {\it et al.} \cite{kumano}) 
nuclear parton distributions. Inside each of the pannels, the two upper curves are obtained with 
I) $ N^{J/\psi}_{th} = 3 \, N^{J/\psi}_{QCD}$  and the two lower curves with 
II) $ N^{J/\psi}_{th} = 0.5 \, N^{J/\psi}_{QCD}$. Finally, solid and dashed lines represent the maximal and the 
minimal contribution from the QGP component Eq. (\ref{npsiqgp}).  Although $f_s = 10^{-2}$ it is easy to see that
choosing a smaller value for the screening factor would move the solid lines to the dashed ones, since they contain
already a very small number of ``in-plasma'' born $J/\psi$'s.

In Fig. 5 we see that there is already a rise of $R_{III}$ at decreasing values of $x_F$, 
but this rise has the shape of a plateau (as anticipated in the introduction), starting from $x_F \simeq 0.01$. 
If the screening is less effective, $f_s = 10^{-2}$, then we find a deviation from the 
plateau at RHIC energies, as in  Figs. 5a and 5b, but do not observe any visible effect at SPS energies, as in 
Figs. 5c and 5d. From these last figures we conclude that in-plasma production can hardly be detected  
at SPS since the $x_F$ distribution is always dominated by the thermal (plateau-like) contribution. On the other 
hand, it should be remarked that if a plateau is found experimentally, this is interesting in itself, being a 
strong evidence in favor of the statistical hadronization model, which implies plasma formation.

Now we come back to the prediction made in \cite{huf}, namely that the initially produced $J/\psi$'s after the 
interaction either with hadronic comovers or with the plasma will disappear around  $x_F \simeq 0$ 
giving place to a dip in this region. 
We can split Eq. (\ref{r2}) into two pieces, the first one containing 
initial production  and the second containing the plasma contribution. In order to include the suppression 
resulting from the final state interactions predicted in \cite{huf}, we replace the first piece   
by a parametrization 
of $R_{AB}(x_F,b=0)$, Eq.~(3) of \cite{huf}, keeping our plasma contribution the same and   construct a new
version of (\ref{r2}): 
\begin{equation}
R_{IV}(x_F) =   R_{AB}(x_F) \, + \,  
\frac{  S^{COM}_{FSI}  \, 
\frac{d N^{AA \rightarrow J/\psi }_{QGP}}{d x_F} }                   
{  T_{AA}(b=0)  \frac{d \sigma^{pp \rightarrow J/\psi }}{d x_F} }
\label{r4}
\end{equation}
This quantity, plotted in Fig. 6 is a more realistic version of $R_{II}$  (plotted in Fig. 4). 
As in previous figures solid and dashed lines correspond to the nuclear parton densities of 
Refs. \cite{eskola1} and \cite{kumano} respectively. Figs. 6a and 6b correspond to RHIC and SPS 
collisions respectively. Following the same convention employed in Fig. 4, in each pannel  the two 
upper curves represent the largest QGP contribution, whereas the two lower curves represent the 
smallest QGP contribution. Comparing Figs. 4 and 6 we can see that the inclusion of FSI changes  
the results, decreasing the strength of the low $x_F$ enhancement. Nevertheless, the effect of
in-plasma production remains quite visible. Notice that the lowest line crosses the origin of the 
horizontal axis ($x_F \simeq 0$) at $0.16$. This is still far above the point predicted in \cite{huf}, 
of $0.055$.  

We discuss now two more conservative scenarios, i.e.,  with less QGP 
production. First, we consider the lower part of 
Fig. 6a, amplifying the scale and allowing for smaller values of the screening factor $f_s$ ($f_s = 10^{-3}$). 
This amplification is shown in Fig. 7a).  
Inspite of the uncertainties in
the calculations, our results in this figure, being still  
larger than the one found in \cite{huf} (the dash-dotted line),  
suggest that an enhancement in
$R_{IV}$ in the low $x_F$  region  might be seen if QGP would be formed. 
Finally, we compute again the QGP production rate, using the same inputs 
except for the initial temperature of the plasma, which we take to be 
$T_0=300$ MeV.  At these smaller temperatures the screening factor must be 
larger and we choose it to be $f_s = 0.1$. This is shown in Fig. 7b) with 
the same conventions used in the previous figures. As it can be seen, our
lowest line is still above the result of  \cite{huf}.

\section{Conclusions}

In this paper we have considered $J/\psi$ production  in nucleus-nucleus collisions by three different mechanisms:
direct (primordial parton fusion)  ii) thermal (statistical coalescence at hadronization)  and 
iii) QGP (in-plasma parton fusion).  As already discussed in \cite{rapp} i) and ii) are just two 
extreme cases. In realistic  simulations, for a fixed number of $c \bar c$ pairs and of $J/\psi$'s, these last 
are gradually destroyed, giving origin to the ``regenerated'' component ii). Component iii) is reconsidered here 
after being forgotten for long time. Our main point was that, even being small, contribution iii) is very strongly 
peaked around $x_F \simeq 0$ and can thus become visible if enough plasma is formed. 

With this in mind we defined the  ratio (\ref{r1}) and computed it assuming mechanism i)  plus a
small component of iii) (ratio $R_{II}$) and assuming mechanism ii) plus a
small component of iii) (ratio $R_{III}$). 
In the first case it is completely flat without plasma contribution. A QGP 
contribution creates a big bump at $x_F < 0.01$.  In the second case, without plasma contribution, we observe
a step structure, with a plateau at $x_F < 0.01$. Switching on the plasma contribution creates a steeply 
falling curve. We believe that any of these features can be measured and will be very interesting for charm 
physics at RHIC. 

In \cite{huf} it was suggested that  the $x_F$ distribution to be measued at RHIC is flat and has a dip in the
region  $x_F \simeq 0$. Our calculations indicate that the inclusion of the 
``in-plasma born'' $J/\psi$'s will fill this dip.

Of course, there are several improvements that one could do in the in-plasma production, as, for example, the use
of a more sophisticated hydrodynamical expansion. Moreover, final state interactions of this component should be 
considered. Work along these lines is already in progress. 
Our findings in this work encourage us to pursue this program.

\vspace{0.5cm}

\underline{Acknowledgements:} This work was partially supported by FAPESP under contract  00/04422-7. We 
are grateful to R. Vogt, P. Levai, X.N. Wang and W. Zajc for useful discussions.

\newpage

\begin{table}[!h]
\begin{tabular}{ccccc}
QCD & \multicolumn{2}{c}{$m_c = 1.2$ \,GeV} & \multicolumn{2}{c}{$m_c = 1.5$ \,GeV} \\ \hline
 & Eskola {\it et al.} \cite{eskola1} & Hirai {\it et al.} \cite{kumano} & Eskola {\it et al.} \cite{eskola1} & Hirai {\it et al.} \cite{kumano} \\ 
\hline
$N_{QCD}^{c\bar c}$ (without shadowing)
& $6.52$ & $4.65$ & $2.65$ & $2.07$ \\
$\,\,\,\,\,\,\,\,\,\,\,\,\,\,\,$ (with shadowing) 
& $6.32$ & $4.36$ & $2.61$ & $1.97$ \\
\hline
$N_{QCD}^{J/\psi}$ (without shadowing) 
& $0.096$ & $0.063$ & $0.019$ & $0.013$ \\
$\,\,\,\,\,\,\,\,\,\,\,\,\,\,\,$ (with shadowing)
& $0.088$ & $0.058$ & $0.018$ & $0.012$ \\
\end{tabular}
\caption{Number of ${J/\psi}$'s and $c\bar c$ pairs produced in $Au-Au$ collisions at RHIC from QCD calculations for  
different values of couplings and charm quark masses.}
\label{tab:thermala}
\end{table}

\begin{table}[!h]
\begin{tabular}{ccccc}
QGP & \multicolumn{2}{c}{$m_c = 1.2$ \,GeV} & \multicolumn{2}{c}{$m_c = 1.5$ \,GeV} \\ \hline
& $\alpha_s(M^2)$ & $\alpha_s(T)$ & $\alpha_s(M^2)$ & $\alpha_s(T)$ \\ 
\hline
\,\,\,\,\,\,\,\,\,\,\,\,\,\,\,\,\,\,\,\,\,\,\,\,\,\,\,Levai {\it el al.} \cite{levai}
& $-\,\,\,\,\,\,-\,$ & $-\,\,\,\,\,(3.7)$ & $-\,\,\,\,\,-\,\,\,$ & $\,-\,\,\,(1.1)$ \\
$N_{QGP}^{c\bar c}$\,\,\,\,\,\,\,\,\,\,Rafelski {\it el al.} \cite{raf}
& $-\,\,\,\,\,\,-\,$ & $-\,\,\,\,\,\,\,\,-\,\,\,\,\,$ & $-\,\,(15)$ & $-\,\,\,\,\,\,-\,\,\,\,$ \\
\,\,\,\,\,\,\,\,\,\,\,\,\,\,\,\,\,\,\,\,\,\,\,\,\,\,\,M\" uller {\it el al.} \cite{lmw}
& $-\,\,\,\,\,\,-\,$ & $-\,\,\,\,\,\,\,\,-\,\,\,\,\,$ & $17\,\,\,\,-\,\,\,$ & $-\,\,\,\,\,\,-\,\,\,\,$ \\
\,\,\,\,\,\,\,\,\,\,\,\,\,\,\,\,\,\,\,\,\,\,\,\,\,\,\,This work
& $120\,(60)$ & $39\,(19.5)$ & $22\,(11)$ & $7.6\,(3.8)$ \\ 
\hline 
$N^{J/\psi}_{QGP}\,\,\,\,\,\,$ This work & $2.70$ & $0.84$ & $0.49$ & $0.16$ \\
\end{tabular}
\caption{Number of ``in-plasma'' produced  $c\bar c$ pairs and ${J/\psi}$'s  in RHIC collisions.   
Numbers inside parenthesis are obtained with $\kappa=1$ and the others with $\kappa=2$. The initial temperature 
is $550$ MeV in all cases.}
\label{tab:thermalb}
\end{table}

\begin{figure}[h]
\centerline{\epsfig{figure=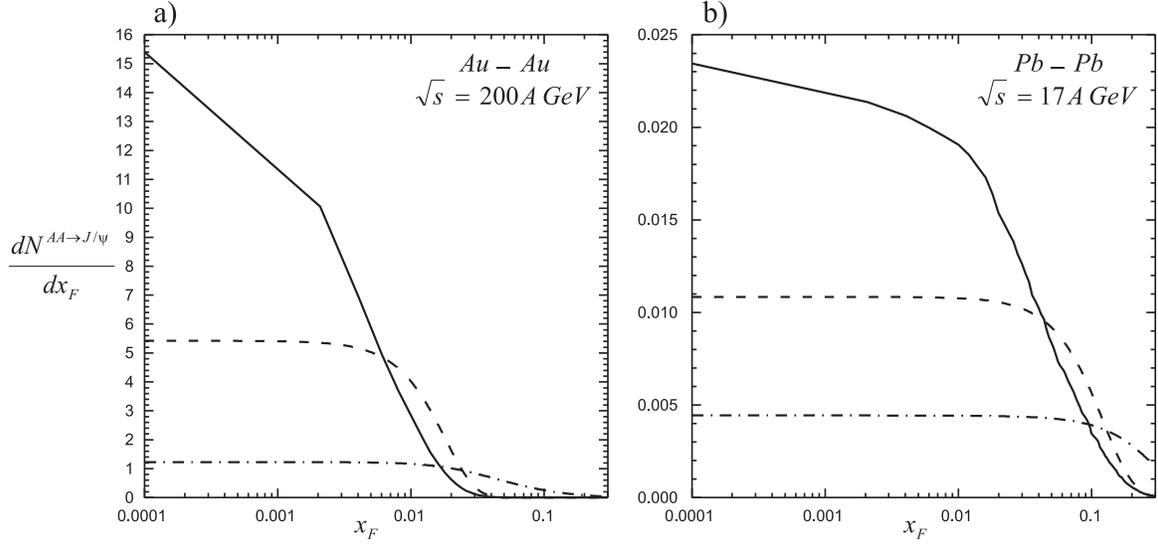,width=15.5cm}}
\caption{$J/\psi$ momentum distributions: direct QCD (\ref{npsiqcd}) (dash-dotted line), in-plasma
 (\ref{npsiqgp}) (solid line) and  thermal production (\ref{dnthermal}) (dashed line). All curves have the 
same normalization (see text). a) RHIC collisions; b) SPS collisions. }
\label{Figure 1ab}
\end{figure}

\begin{figure}[h]
\centerline{\epsfig{figure=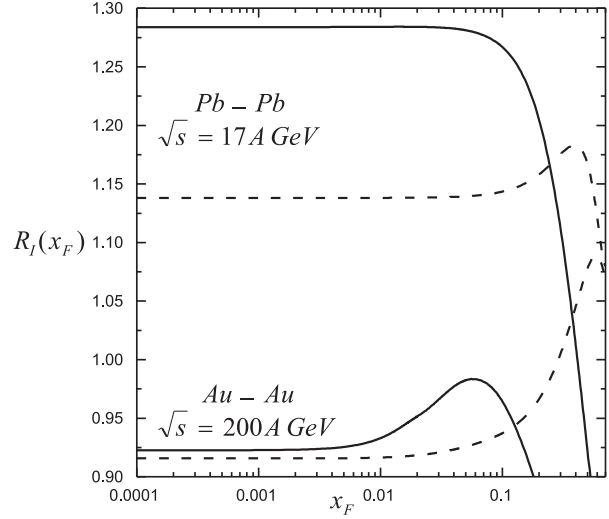,width=8.cm}}
\caption{Ratio $R_I(x_F)$: effect of shadowing in QCD direct production. The numerators were obtained 
with (\ref{npsiqcd}) and (\ref{cem5})   and the nuclear parton distribution functions were taken from  
\protect\cite{eskola1} (solid lines)  and  from \protect\cite{kumano} (dashed lines).}
\label{Figure 2}
\end{figure}

\begin{figure}[h]
\centerline{\epsfig{figure=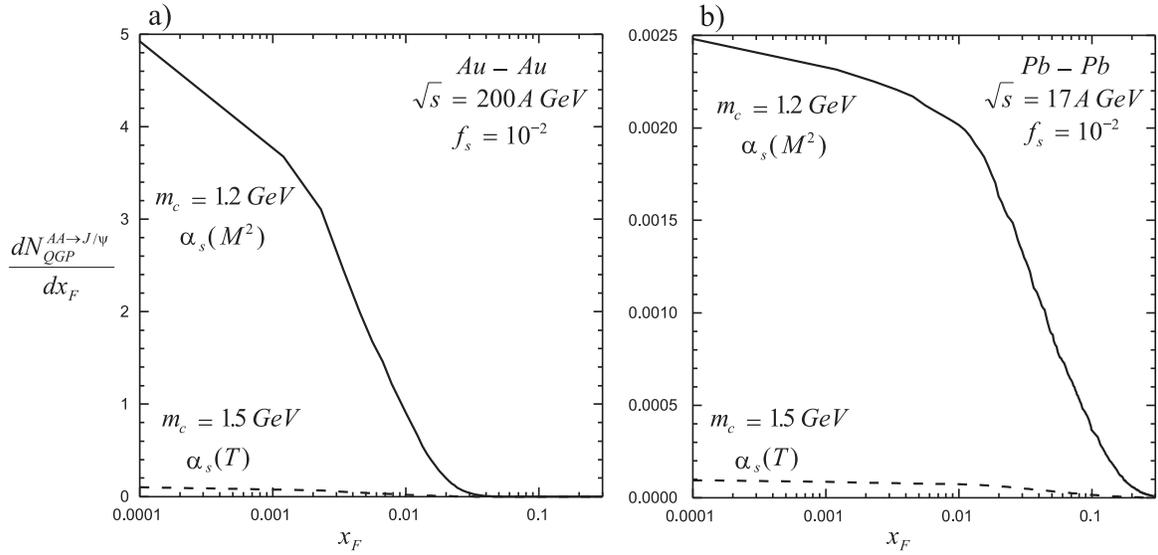,width=15.5cm}}
\caption{In-plasma $J/\psi$ momentum distribution: a)  $Au+Au$ at RHIC; b)  $Pb+Pb$ at SPS. Solid (dashed) 
lines are obtained with the smallest (largest) charm quark mass and with $\alpha (M^2)$ ($\alpha (T)$). The 
curves are obtained with expression (\ref{npsiqgp}) and in both cases the screening factor ($f_s$) has been 
fixed to $10^{-2}$.}
\label{Figure 3ab}
\end{figure}

\begin{figure}[h]
\centerline{\epsfig{figure=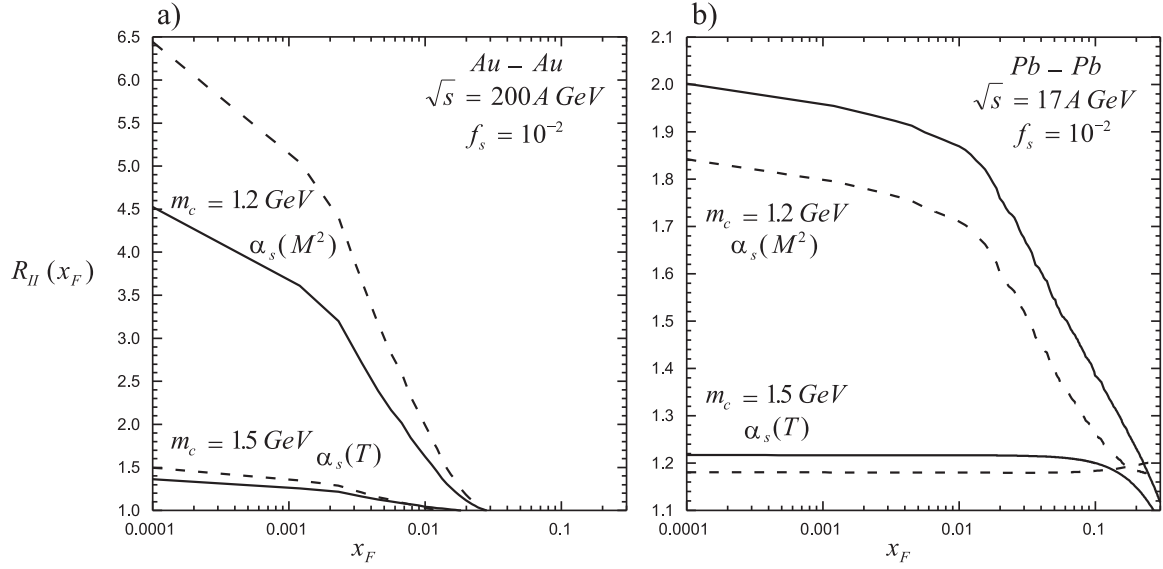,width=15.5cm}}
\caption{Ratio $R_{II}(x_F)$: a)  $Au+Au$ at RHIC; b)  $Pb+Pb$ at SPS. Solid and dashed lines, as in Fig. 2, are 
obtained with the nuclear parton densities taken from \protect\cite{eskola1} and \protect\cite{kumano} 
respectively.}
\label{Figure 4ab}
\end{figure}

\begin{figure}[h]
\centerline{\epsfig{figure=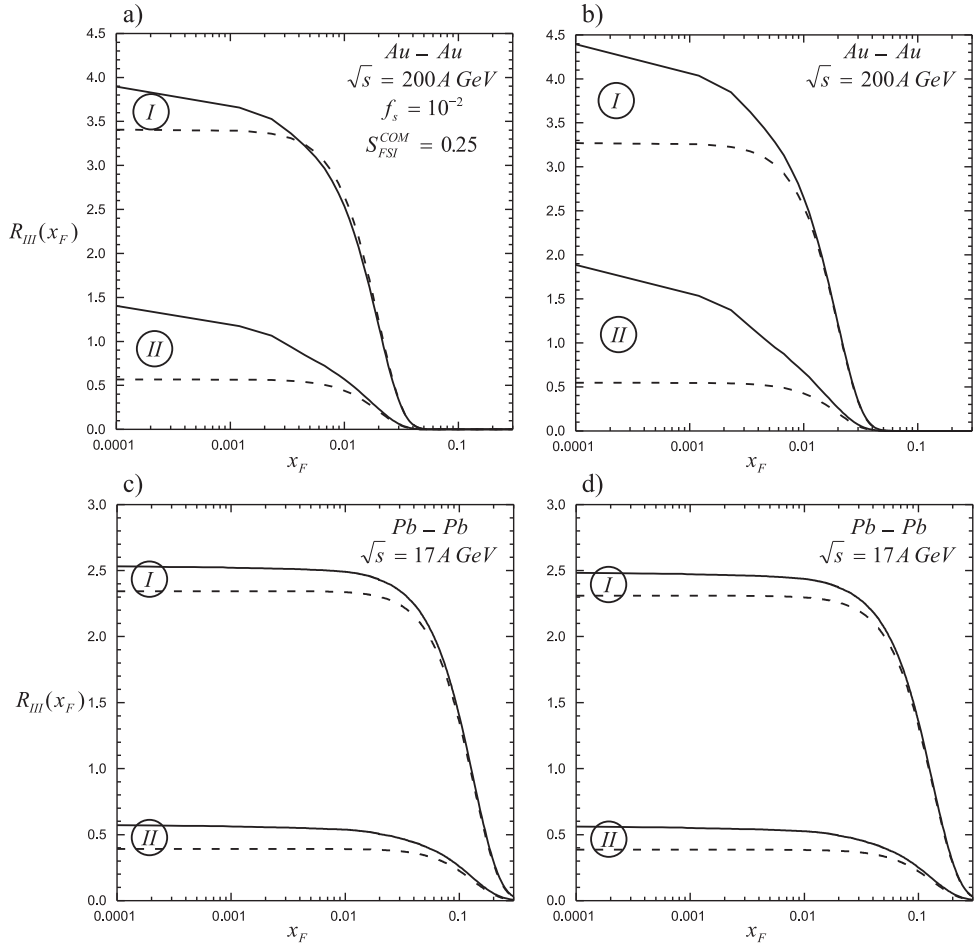,width=13.0cm}}
\caption{Ratio $R_{III}(x_F)$ calculated with (\ref{r3}); labels I and II represent the normalizations
defined in (\ref{normal}); solid (dashed) lines are obtained with the largest (smallest) in-plasma 
contribution; a) and b) are for RHIC collisions and c) and d) for SPS collisions; in the two left (right) pannels 
we use the Eskola {\it et al.} \protect\cite{eskola1}  (Hirai {\it et al.} \protect\cite{kumano}) nuclear parton 
distributions.}
\label{Figure 5abcd}
\end{figure}

\begin{figure}[h]
\centerline{\epsfig{figure=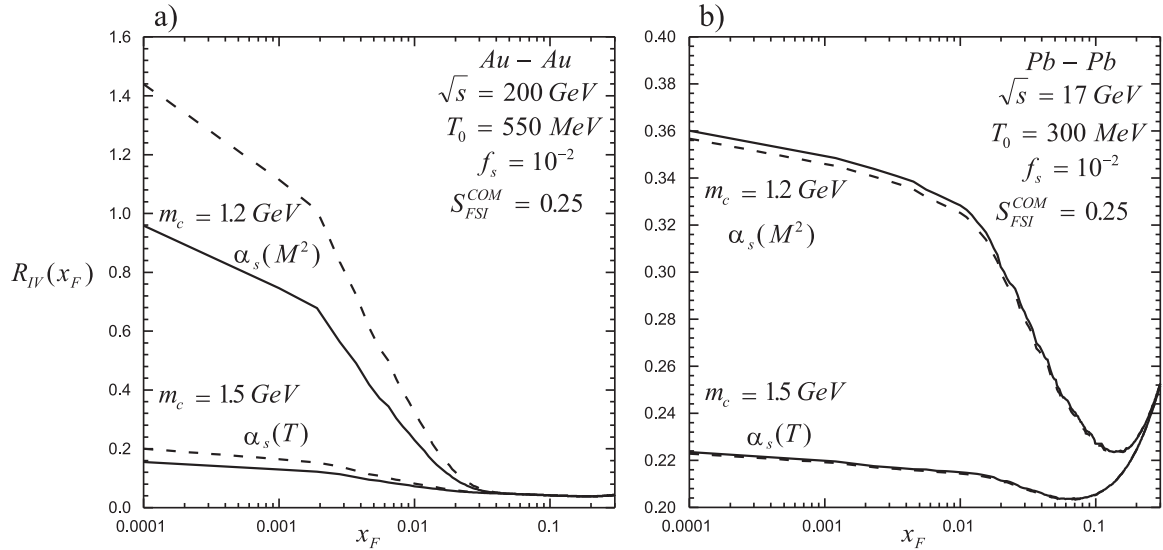,width=15.5cm}}
\caption{Ratio $R_{IV}(x_F)$ computed with (\ref{r4}). a) RHIC collisions; b) SPS collisions; solid lines:
parton densities of Ref. \protect\cite{eskola1}; dashed lines: parton densities of Ref. \protect\cite{kumano}}
\label{Figure 6ab}
\end{figure}

\begin{figure}[h]
\centerline{\epsfig{figure=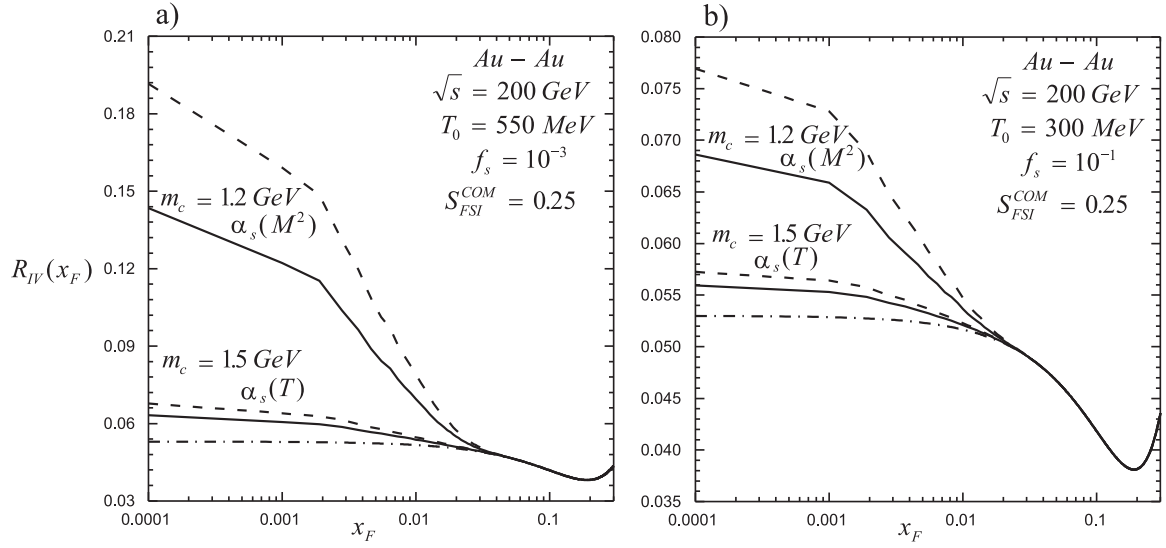,width=15.5cm}}
\caption{Ratio $R_{IV}(x_F)$ computed with (\ref{r4})  for RHIC collisions; solid lines:
parton densities of Ref. \protect\cite{eskola1}; dashed lines: parton densities of Ref. \protect\cite{kumano}; dash-dotted line represents the (parameterized) result for central collisions obtained by H\"ufner {\it et al.} \protect\cite{huf}. 7a) $T_0 = 550$ MeV; 7b) $T_0 = 300$ MeV. }
\label{Figure 7}
\end{figure}

\end{document}